\documentclass[aps,pra,reprint,showpacs,showkeys,superscriptaddress]{revtex4-1}

\usepackage{amsmath}
\usepackage{amsfonts}
\usepackage{stmaryrd}
\usepackage{amssymb}
\usepackage{graphics,epsfig}
\usepackage{subfigure}
\usepackage{bm}
\usepackage{mathpple}
\usepackage{CJK}

\begin{document}
\begin{CJK*}{GBK}{}

\title{Could the Earth's surface Ultraviolet irradiance be blamed for the global warming?(I)\\
        \textit{------A new effect may exist}}

\author{Jilong Chen}
\affiliation{Department of Astronomy, Beijing Normal University,Beijing 100875, China}
\email{niba180634@gmail.com}

\author{Zhaopeng Sun}
\affiliation{School of Physics, Shandong University, Jinan 250100, China}

\author{Juan Zhao}
\affiliation{Department of Astronomy, Beijing Normal University,Beijing 100875, China}

\author{Yujun Zheng}
\email{yzheng@sdu.edu.cn}
\affiliation{School of Physics, Shandong University, Jinan 250100, China}

\begin{abstract}
 Whether natural factors could interpret the rise of the Earth's surface
 temperature is still controversial. Though numerous recent researches have
 reported apparent correlations between solar activity and the Earth's climate,
 solar activity has encountered a big problem when describing the rapid
 global warming after 1970s. Our investigation shows the good positive
 correlations between the Earth's surface Ultraviolet irradiance (280-400 nm)
 and the Earth's surface temperature both in temporal and spatial
 variations by analyzing the global surface Ultraviolet irradiance (280-400 nm)
 and global surface temperature data from 1980-1999. The rise of CO$_2$
 cannot interpret the good positive correlations, and we could even get an
 opposite result to the good correlations when employing the rise of CO$_2$
 to describe the relation between them. Based on the good positive correlations,
 we suggest a new effect, named ``Highly Excited Water Vapor" (HEWV) effect,
 which can interpret how the Sun influences the Earth's surface temperature
 reasonably, including the rapid warming after 1970s.
\end{abstract}



\maketitle

\section{introduction}
\label{sec:int}

The debate on whether the Sun could influence global surface temperature variation
has been argued for years since 1976 after Eddy \cite{eddy1976} indicated the relation
between solar activity and Earth's surface temperature. Many researchers have
found apparent correlations between solar activity and the Earth's surface temperature \cite{reid1987,friis1991,lassen1995,lean1995,solanki1999} and some physical
mechanisms have been suggested to interpret the possible way of solar activity
influencing global surface temperature \cite{cubasch2000,haigh1994,haigh1996,Reid2000,haigh2010,Svensmark1997}.
Some researchers, however, have also found a decoupling between solar activity and the
Earth's surface temperature since roughly 1970: the surface temperature continued
to rise rapidly, while the solar irradiance did not show a corresponding increase in the same time \cite{Thejll,solanki1998,Lockwood2008}. Solanki and Krivova \cite{Lockwood2008} compared the
constructed total solar irradiance, UV irradiance and the cosmic rays with the atmosphere
temperature respectively, and got a conclusion that since roughly 1970
he Sun cannot have contributed more than $30\%$ to the steep temperature increase.

We consider the decoupling problem between solar activity and the Earth's
surface temperature being caused by the decrease of the Earth's ozone layer and
cloud cover. As we all know, solar Ultraviolet irradiance could be influenced by
ozone layer\cite{cutchis1974} and the cloud cover\cite{Josep2005}.
o we employ the Earth's surface Ultraviolet irradiance (280-400 nm, the same hereinafter)
as the index of their variations, and via comparing it with the surface temperature
to investigate the relation between surface Ultraviolet irradiance and surface
temperature. We could see good positive correlations between surface temperature
and surface Ultraviolet irradiance in Sec.~\ref{sec:result}. And the physical mechanics behind
the good correlations is discussed in Sec.~\ref{sec:dis}, one probable mechanics is the
cloud cover, and another, is the new effect raised by Chen {it et al.},
named as ``Highly Excited Water Vapor" effect.

\section{data}
\label{sec:theory}

Climatological distributions of the Earth's surface-level Ultraviolet
radiation data from 1980 to 1999 is obtained from \cite{julia2010},
and this data is the monthly mean data, on $60.5 \rm^o$S-$60.5 \rm^o$N,
$179.5 \rm^o$W-$179.5 \rm^o$E ($\rm 1^o\times1.25^o$) from 1980 to 1999,
divided into several parts based on the spectra band, such as UVA: 325-400 nm, UVB: 280-325 nm.
Then we interpolated and averaged the monthly data to the yearly anomalies
in the spectra band 280-400 nm (UVA data plus UVB data) in $\rm 1^o\times1^o$.
The yearly surface temperature anomalies on $60.5 \rm^o$S-$60.5 \rm^o$N,
$179.5 \rm^o$W-$179.5 \rm^o$E ($\rm 1^o\times1^o$) from 1980 to 1999 are
downloaded directly from Goddard Institute for Space Studies (GISS),
National Aeronautics and Space Administration (NASA).

\section{results}
\label{sec:result}

\begin{figure}[h]
   \includegraphics[width=\hsize]{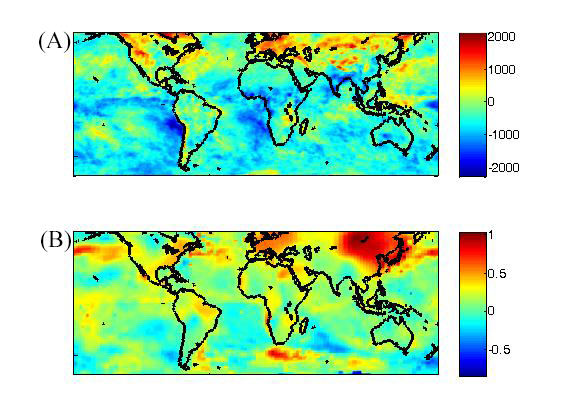}
   \caption{Comparison of surface temperature distribution and surface Ultraviolet
   irradiance distribution in decadal variation (average of 1990-1999 minus average of 1980-1989).
   (A): Earth's surface Ultraviolet irradiance (unit: $\rm kJ.m^{-2}.month^{-1}$);
   (B): Earth's surface temperature (unit: $\rm^oC$).}
   \label{Fig1}%
\end{figure}

By taking investigation of the two datasets, we find the good correlation in spatial
distribution ($r=0.3855$, $P>99.99\%$) between surface Ultraviolet irradiance and
surface temperature in decadal variation. It is shown in Fig.~\ref{Fig1}.
Surface Ultraviolet irradiance and surface temperature show large increase
in latitude zone $30\rm^o$N-$60\rm^o$N, especially in Siberia, West Europe and North America.

\begin{figure}[h]
   \includegraphics[width=\hsize]{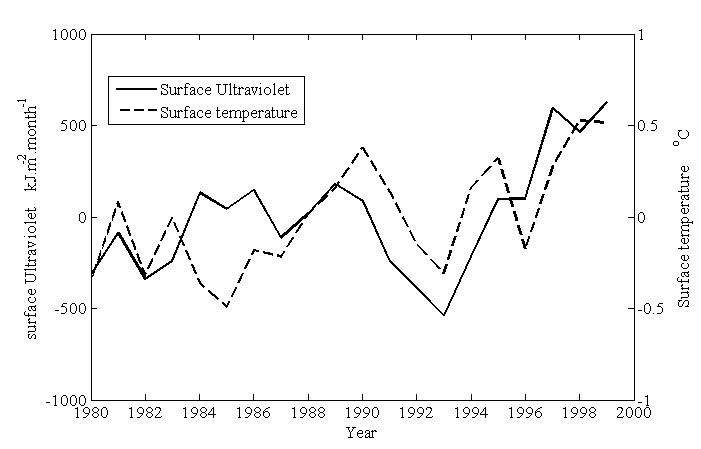}
   \caption{The 30$\rm ^o$N-60$\rm ^o$N yearly zonal mean variation of Earth's surface
   Ultraviolet irradiance (the solid line) and Earth's surface temperature (the dashed line) from 1980 to 1999.}
   \label{Fig2}%
    \end{figure}

Because of the consistent largest increase between surface Ultraviolet irradiance and
surface temperature in $30 \rm^o$N-$60 \rm^o$N, yearly zonal mean variation of them in this
latitude region from 1980 to 1999 are calculated and compared. The result is shown in Fig.~\ref{Fig2}.
We find that the surface Ultraviolet irradiance correlates with surface Ultraviolet irradiance
very well ($r=0.6022$, $P>99.5\%$), except the years 1983-1984, 1989-1990 and 1997-1999,
around El Ni\~{n}o appearance.

\section{Discussion and suggestion}
\label{sec:dis}

Because the Earth's surface temperature is not considered in the process of
the Earth's surface ultraviolet irradiance calculation, and vice versa,
so the good correlations are not caused by artificial calculation.
That is, the good correlations indicate an exact physical relation between
surface Ultraviolet irradiance and surface temperature. The rise of $\rm{CO} _2$
cannot be employed to describe the good correlations. Instead, it presents an
opposite result, because based on the substantial hypothesis that in the
warmer and moister CO$\rm _2$-rich atmosphere, cloud liquid water content
will generally be larger and will increase the cloud's albedo \cite{Somerville1984},
which will lead a decrease of the Ultraviolet irradiance. This leads to an
opposite result to the good positive correlations.

\subsection{Could the cloud cover cause the good correlations?}

Cloud cover, as a natural factor, may be a common factor to the good
positive correlations between the Earth's surface temperature and surface
Ultraviolet irradiance, because cloud cover has influence on both
ltraviolet irradiance \cite{julia2010} and global surface temperature \cite{Schneider1972,hartmann1994}.
The decrease of the cloud could lead the increase of surface temperature and
Ultraviolet irradiance at the same time, and cause the similarity between surface
Ultraviolet irradiance and surface temperature. The International Satellite Cloud
Climatology Project (ISCCP) did find the declining of the total cloud cover \cite{Schiffer1983},
which gives us a positive motivation to research the effect of the cloud on
the surface temperature variation.

\begin{figure}[h]
   \includegraphics[width=\hsize]{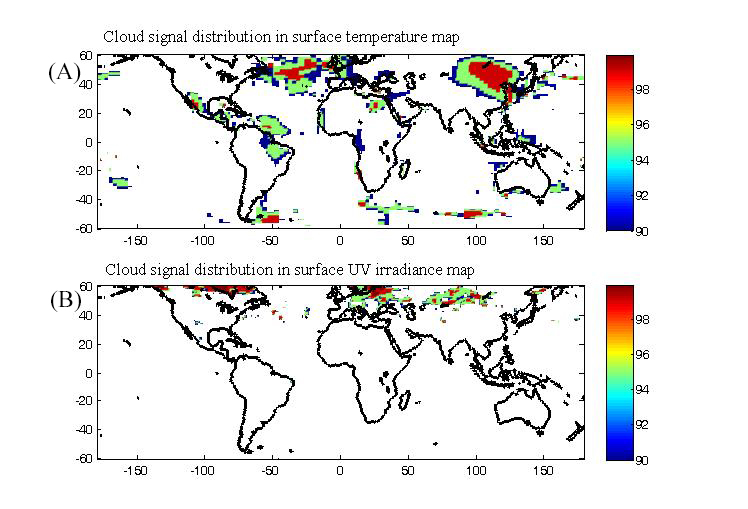}
   \caption{Spatial distribution of surface temperature
   (A) and surface Ultraviolet irradiance
   (B) responding to total cloud coverage signal above $90\%$ confidence level.
   (Color bar: confidence level)}
   \label{Fig3}%
\end{figure}

We test the cloud signal in the Earth's surface temperature data and the
Earth's surface Ultraviolet irradiance data, respectively, by employing
the total cloud coverage yearly variation from 1983 to 1999 \cite{Schiffer1983}.
We find both of the surface temperature and surface Ultraviolet irradiance
data show their responses to the total cloud coverage signal,
but the areas in the two data maps showing their responses to the
cloud signal are very different (see Fig.~\ref{Fig3}). This could indicate that
the cloud is not the main factor caused the good positive correlations
between surface temperature and surface Ultraviolet irradiance,
but more researches need to be done before we get the final conclusion.

\subsection{A new effect may exist}

The good correlations between surface Ultraviolet irradiance and surface
temperature could indicate a new effect: the surface Ultraviolet irradiance
could have the ability to influence surface temperature variation directly.
As we know, water vapor is the most important greenhouse gas, accounting
for almost $70\%$ of the greenhouse effect \cite{Freidenreich1993}.
Any change of it could lead a surface temperature change. Actually,
water vapor has many properties that deserve to be paid more attention \cite{Bernath2002},
specifically its effects in highly excited vibrational states \cite{Zheng2000}.
Here we concentrate on the effects of water vapor in highly vibrational states
excited by Ultraviolet irradiance and propose a new suggestion as one of
the paths that solar activity influences the Earth's surface temperature variation.

\begin{figure}[h]
   \includegraphics[width=\hsize]{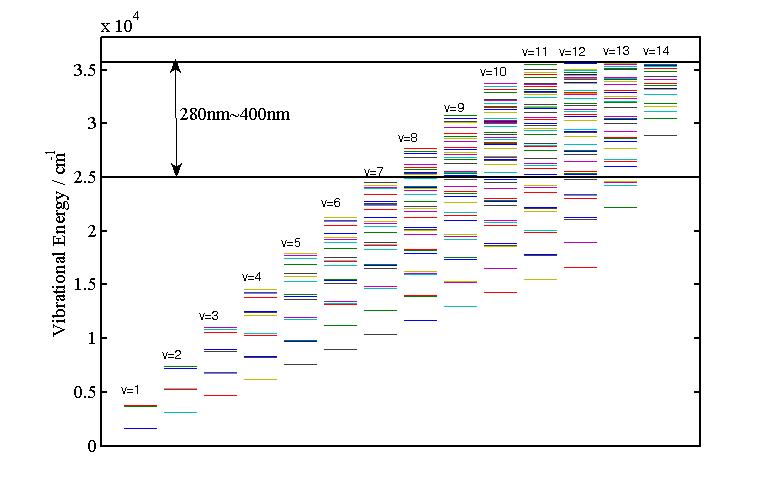}
   \caption{Vibrational energy states of water molecule.
   The state region between the two bold-lines is the high
   vibrational energy state region excited by the Ultraviolet
   irradiance in spectral band 280-400 nm.}
   \label{Fig4}%
\end{figure}

Figure~\ref{Fig4} shows the vibrational energy states of a water molecule,
which is calculated by Lie algebraic approach \cite{Zheng2000,Iachello1995}.
Water molecule at the low vibrational states can be excited by the Ultraviolet
irradiance (280-400 nm) to a highly excited vibrational state region between
the two bold-lines marked in Fig.~\ref{Fig4}, where there are more vibrational
energy states and generate many new absorption lines and strengthen
some absorption lines in infrared spectrum region.

\begin{figure}[h]
   \includegraphics[width=\hsize]{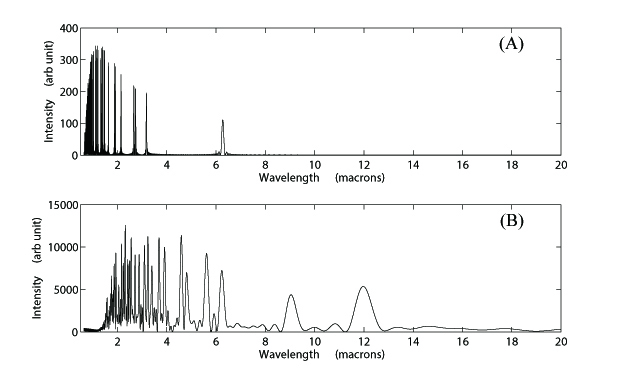}
   \caption{Absorption spectrum comparison of the low vibrational energy water molecule
   (A) and highly excited water molecule excited by 380 nm Ultraviolet irradiance (B).}
   \label{Fig5}%
\end{figure}

The water molecular spectroscopy calculation indicates that Ultraviolet
irradiance changes water vapor's infrared spectrum. The Infrared spectrum
of water vapor in the highly excited vibrational states excited by the 380 nm Ultraviolet
irradiance is shown in Fig.~\ref{Fig5}(B). By taking comparison with water vapor's
Infrared spectrum in low vibrational states (Fig.~\ref{Fig5}(A)),
the spectra of water vapor in highly excited vibrational states shows
many new absorption lines, such as absorption lines in $4-6~\mu m$,
$9~\mu m$ and $12~\mu m$. The $9~\mu m$ and $12~\mu m$
absorption lines are in the Earth's $8-14~\mu m$ atmospheric window,
and the atmospheric window means this part of spectrum dose not be
absorbed by the Earth's atmosphere and just get through to the outer space.
Also these two lines are in the spectra region that the Earth's surface
emits its maximum energy as a black body when its surface temperature
is suggested as 288 K. This means we find more energy source for water
vapor's greenhouse effect, in other words, for the global warming.

\begin{figure}[h]
   \includegraphics[width=\hsize]{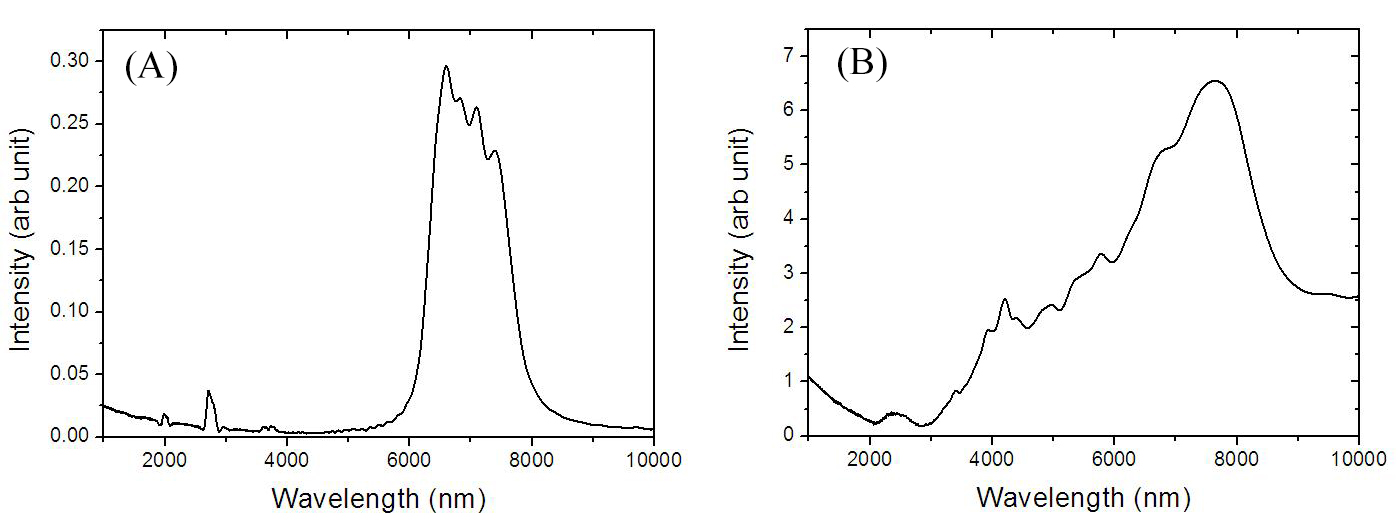}
   \caption{Comparison of water molecule's photon absorption cross
   section in low vibrational energy states (A) and in highly excited
   vibrational states excited the 380 nm Ultraviolet irradiance (B)}
    \label{Fig6}%
\end{figure}

We calculate the highly excited water molecule's photon absorption cross
section, which is excited by the 380nm Ultraviolet irradiance, and
compared with the low vibrational water molecule's photon absorption
cross section. We find that the photon absorption cross section of
the highly excited water molecule is almost 20-100 times larger than
that of the low energy state (see Fig.~\ref{Fig6}), which means the absorption
ability of highly excited water molecule is much stronger than that
of low vibrational energy water molecule.

The Ultraviolet irradiance in 380 nm is just one point of the 280-400 nm
spectra band. We could also employ Ultraviolet irradiance in other wavelength
to calculate highly excited water molecular spectroscopy, and could get
other new absorption lines in water molecular and find more energy resource for the global warming.

\begin{figure}[]
   \includegraphics[width=\hsize]{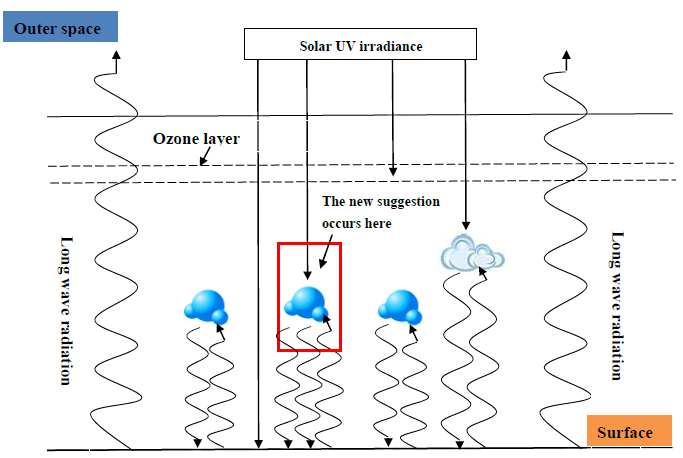}
   \caption{The physical process that the HEWV effect occurs in the Earth-atmosphere system}
   \label{Fig7}%
\end{figure}

Based on the above analysis, we suggest that the Ultraviolet irradiance
in the spectral band 280-400 nm in the Earth's lower troposphere can
be absorbed by water vapor, and then change water vapor's infrared
absorption spectrum, generating many new absorption lines. In other words,
the Earth's surface Ultraviolet irradiance enhances water vapor's greenhouse
effect and eventually influences the Earth's surface temperature.
The chart description of this physical process is shown in Fig.~\ref{Fig7}.
We name this as ``Highly Excited Water Vapor" (HEWV) effect.

The HEWV effect can highlight the importance of the solar activity in
global surface temperature variation. Solar activity has the ability to
influence global surface temperature variation through Ultraviolet
irradiance in spectral band 280-400 nm. The role of ozone layer and
cloud in this effect is like a thermostat, which absorbs or reflects
Ultraviolet irradiance and tries to weaken the ability of solar activity
influencing global surface temperature variation. After 1970s,
the ozone layer and the cloud cover decrease, which makes the increase
of the Earth's surface Ultraviolet radiance and eventually leads the
increase of global surface mean temperature.

\begin{acknowledgements}
This work is supported by National Natural Science Foundation of
China (Grant Nos. 11203004, 10978007, 11374191 and 91021009).
The surface temperature data is from Goddard Institute for Space Studies (GISS),
National Aeronautics and Space Administration (NASA). The total cloud coverage
data is from the International Satellite Cloud Climatology Project (ISCCP)
We are grateful to Julia Lee-Taylor for providing the surface Ultraviolet
irradiance data, to Zi-Niu Xiao for discussion.
\end{acknowledgements}

\end{CJK*}


\begin{thebibliography}{26}
\expandafter\ifx\csname natexlab\endcsname\relax\def\natexlab#1{#1}\fi
\expandafter\ifx\csname bibnamefont\endcsname\relax
  \def\bibnamefont#1{#1}\fi
\expandafter\ifx\csname bibfnamefont\endcsname\relax
  \def\bibfnamefont#1{#1}\fi
\expandafter\ifx\csname citenamefont\endcsname\relax
  \def\citenamefont#1{#1}\fi
\expandafter\ifx\csname url\endcsname\relax
  \def\url#1{\texttt{#1}}\fi
\expandafter\ifx\csname urlprefix\endcsname\relax\def\urlprefix{URL }\fi
\providecommand{\bibinfo}[2]{#2}
\providecommand{\eprint}[2][]{\url{#2}}

\bibitem[{\citenamefont{Eddy}(1976)}]{eddy1976}
\bibinfo{author}{\bibfnamefont{J.~A.} \bibnamefont{Eddy}},
  \bibinfo{journal}{Science} \textbf{\bibinfo{volume}{192}},
  \bibinfo{pages}{1189} (\bibinfo{year}{1976}).

\bibitem[{\citenamefont{Reid}(1987)}]{reid1987}
\bibinfo{author}{\bibfnamefont{G.~C.} \bibnamefont{Reid}},
  \bibinfo{journal}{Nature} \textbf{\bibinfo{volume}{329}},
  \bibinfo{pages}{142} (\bibinfo{year}{1987}).

\bibitem[{\citenamefont{Friis-Christensen and Lassen}(1991)}]{friis1991}
\bibinfo{author}{\bibfnamefont{E.}~\bibnamefont{Friis-Christensen}}
  \bibnamefont{and} \bibinfo{author}{\bibfnamefont{K.}~\bibnamefont{Lassen}},
  \bibinfo{journal}{Science} \textbf{\bibinfo{volume}{254}},
  \bibinfo{pages}{698} (\bibinfo{year}{1991}).

\bibitem[{\citenamefont{Lassen and Friis-Christensen}(1995)}]{lassen1995}
\bibinfo{author}{\bibfnamefont{K.}~\bibnamefont{Lassen}} \bibnamefont{and}
  \bibinfo{author}{\bibfnamefont{E.}~\bibnamefont{Friis-Christensen}},
  \bibinfo{journal}{Journal of Atmospheric and Terrestrial Physics}
  \textbf{\bibinfo{volume}{57}}, \bibinfo{pages}{835} (\bibinfo{year}{1995}).

\bibitem[{\citenamefont{Lean et~al.}(1995)\citenamefont{Lean, Beer, and
  Bradley}}]{lean1995}
\bibinfo{author}{\bibfnamefont{J.}~\bibnamefont{Lean}},
  \bibinfo{author}{\bibfnamefont{J.}~\bibnamefont{Beer}}, \bibnamefont{and}
  \bibinfo{author}{\bibfnamefont{R.}~\bibnamefont{Bradley}},
  \bibinfo{journal}{Geophysical Research Letters}
  \textbf{\bibinfo{volume}{22}}, \bibinfo{pages}{3195} (\bibinfo{year}{1995}).

\bibitem[{\citenamefont{Solanki and Fligge}(1999)}]{solanki1999}
\bibinfo{author}{\bibfnamefont{S.~K.} \bibnamefont{Solanki}} \bibnamefont{and}
  \bibinfo{author}{\bibfnamefont{M.}~\bibnamefont{Fligge}},
  \bibinfo{journal}{Geophysical Research Letters}
  \textbf{\bibinfo{volume}{26}}, \bibinfo{pages}{2465} (\bibinfo{year}{1999}).

\bibitem[{\citenamefont{Cubasch and Voss}(2000)}]{cubasch2000}
\bibinfo{author}{\bibfnamefont{U.}~\bibnamefont{Cubasch}} \bibnamefont{and}
  \bibinfo{author}{\bibfnamefont{R.}~\bibnamefont{Voss}},
  \bibinfo{journal}{Space Science Reviews} \textbf{\bibinfo{volume}{94}},
  \bibinfo{pages}{185} (\bibinfo{year}{2000}).

\bibitem[{\citenamefont{Haigh}(1994)}]{haigh1994}
\bibinfo{author}{\bibfnamefont{J.~D.} \bibnamefont{Haigh}},
  \bibinfo{journal}{Nature} \textbf{\bibinfo{volume}{370}},
  \bibinfo{pages}{544} (\bibinfo{year}{1994}).

\bibitem[{\citenamefont{Haigh}(1996)}]{haigh1996}
\bibinfo{author}{\bibfnamefont{J.~D.} \bibnamefont{Haigh}},
  \bibinfo{journal}{Science} \textbf{\bibinfo{volume}{272}},
  \bibinfo{pages}{981} (\bibinfo{year}{1996}).

\bibitem[{\citenamefont{Reid}(2000)}]{Reid2000}
\bibinfo{author}{\bibfnamefont{G.~C.} \bibnamefont{Reid}},
  \bibinfo{journal}{Space Science Reviews} \textbf{\bibinfo{volume}{94}},
  \bibinfo{pages}{1} (\bibinfo{year}{2000}), ISSN \bibinfo{issn}{0038-6308}.

\bibitem[{\citenamefont{Haigh et~al.}(1996)\citenamefont{Haigh, Winning, Toumi,
  and Harder}}]{haigh2010}
\bibinfo{author}{\bibfnamefont{J.~D.} \bibnamefont{Haigh}},
  \bibinfo{author}{\bibfnamefont{A.~R.} \bibnamefont{Winning}},
  \bibinfo{author}{\bibfnamefont{R.}~\bibnamefont{Toumi}}, \bibnamefont{and}
  \bibinfo{author}{\bibfnamefont{J.~W.} \bibnamefont{Harder}},
  \bibinfo{journal}{Science} \textbf{\bibinfo{volume}{272}},
  \bibinfo{pages}{981} (\bibinfo{year}{1996}).

\bibitem[{\citenamefont{Svensmark and Friis-Christensen}(1997)}]{Svensmark1997}
\bibinfo{author}{\bibfnamefont{H.}~\bibnamefont{Svensmark}} \bibnamefont{and}
  \bibinfo{author}{\bibfnamefont{E.}~\bibnamefont{Friis-Christensen}},
  \bibinfo{journal}{Journal of Atmospheric and Solar-Terrestrial Physics}
  \textbf{\bibinfo{volume}{59}}, \bibinfo{pages}{1225 } (\bibinfo{year}{1997}).

\bibitem[{\citenamefont{Thejll and Lassen}(2000)}]{Thejll}
\bibinfo{author}{\bibfnamefont{P.}~\bibnamefont{Thejll}} \bibnamefont{and}
  \bibinfo{author}{\bibfnamefont{K.}~\bibnamefont{Lassen}},
  \bibinfo{journal}{Journal of Atmospheric and Solar-Terrestrial Physics}
  \textbf{\bibinfo{volume}{62}}, \bibinfo{pages}{1207 } (\bibinfo{year}{2000}).

\bibitem[{\citenamefont{Solanki and Fligge}(1998)}]{solanki1998}
\bibinfo{author}{\bibfnamefont{S.~K.} \bibnamefont{Solanki}} \bibnamefont{and}
  \bibinfo{author}{\bibfnamefont{M.}~\bibnamefont{Fligge}},
  \bibinfo{journal}{Geophysical Research Letters}
  \textbf{\bibinfo{volume}{25}}, \bibinfo{pages}{341} (\bibinfo{year}{1998}).

\bibitem[{\citenamefont{Lockwood and Fr?hlich}(2008)}]{Lockwood2008}
\bibinfo{author}{\bibfnamefont{M.}~\bibnamefont{Lockwood}} \bibnamefont{and}
  \bibinfo{author}{\bibfnamefont{C.}~\bibnamefont{Fr?hlich}},
  \bibinfo{journal}{Proceedings of the Royal Society A: Mathematical, Physical
  and Engineering Science} \textbf{\bibinfo{volume}{464}},
  \bibinfo{pages}{1367} (\bibinfo{year}{2008}).

\bibitem[{\citenamefont{Cutchis}(1974)}]{cutchis1974}
\bibinfo{author}{\bibfnamefont{P.}~\bibnamefont{Cutchis}},
  \bibinfo{journal}{Science} \textbf{\bibinfo{volume}{184}},
  \bibinfo{pages}{13} (\bibinfo{year}{1974}).

\bibitem[{\citenamefont{Calbo et~al.}(2005)\citenamefont{Calbo, Pages, and
  Gonzalez}}]{Josep2005}
\bibinfo{author}{\bibfnamefont{J.}~\bibnamefont{Calbo}},
  \bibinfo{author}{\bibfnamefont{D.}~\bibnamefont{Pages}}, \bibnamefont{and}
  \bibinfo{author}{\bibfnamefont{J.-A.} \bibnamefont{Gonzalez}},
  \bibinfo{journal}{Reviews of Geophysics} \textbf{\bibinfo{volume}{43}},
  \bibinfo{pages}{n/a} (\bibinfo{year}{2005}), ISSN \bibinfo{issn}{1944-9208},
  \urlprefix\url{http://dx.doi.org/10.1029/2004RG000155}.

\bibitem[{\citenamefont{Lee-Taylor et~al.}(2010)\citenamefont{Lee-Taylor,
  Madronich, Fischer, and Mayer}}]{julia2010}
\bibinfo{author}{\bibfnamefont{J.}~\bibnamefont{Lee-Taylor}},
  \bibinfo{author}{\bibfnamefont{S.}~\bibnamefont{Madronich}},
  \bibinfo{author}{\bibfnamefont{C.}~\bibnamefont{Fischer}}, \bibnamefont{and}
  \bibinfo{author}{\bibfnamefont{B.}~\bibnamefont{Mayer}}, in
  \emph{\bibinfo{booktitle}{UV Radiation in Global Climate Change}}, edited by
  \bibinfo{editor}{\bibfnamefont{W.}~\bibnamefont{Gao}},
  \bibinfo{editor}{\bibfnamefont{J.}~\bibnamefont{Slusser}}, \bibnamefont{and}
  \bibinfo{editor}{\bibfnamefont{D.}~\bibnamefont{Schmoldt}}
  (\bibinfo{publisher}{Springer Berlin Heidelberg}, \bibinfo{year}{2010}), pp.
  \bibinfo{pages}{1--20}, ISBN \bibinfo{isbn}{978-3-642-03312-4},
  \urlprefix\url{http://dx.doi.org/10.1007/978-3-642-03313-1_1}.

\bibitem[{\citenamefont{Somerville and Remer}(1984)}]{Somerville1984}
\bibinfo{author}{\bibfnamefont{R.~C.~J.} \bibnamefont{Somerville}}
  \bibnamefont{and} \bibinfo{author}{\bibfnamefont{L.~A.} \bibnamefont{Remer}},
  \bibinfo{journal}{Journal of Geophysical Research: Atmospheres}
  \textbf{\bibinfo{volume}{89}}, \bibinfo{pages}{9668} (\bibinfo{year}{1984}),
  ISSN \bibinfo{issn}{2156-2202},
  \urlprefix\url{http://dx.doi.org/10.1029/JD089iD06p09668}.

\bibitem[{\citenamefont{Schneider}(1972)}]{Schneider1972}
\bibinfo{author}{\bibfnamefont{S.~H.} \bibnamefont{Schneider}},
  \bibinfo{journal}{J. Atmos. Sci.} \textbf{\bibinfo{volume}{29}},
  \bibinfo{pages}{1413} (\bibinfo{year}{1972}).

\bibitem[{\citenamefont{Hartmann}(1994)}]{hartmann1994}
\bibinfo{author}{\bibfnamefont{D.}~\bibnamefont{Hartmann}},
  \emph{\bibinfo{title}{Global Physical Climatology}}, International Geophysics
  (\bibinfo{publisher}{Elsevier Science}, \bibinfo{year}{1994}), ISBN
  \bibinfo{isbn}{9780080571638}.

\bibitem[{\citenamefont{Schiffer and Rossow}(1983)}]{Schiffer1983}
\bibinfo{author}{\bibfnamefont{R.}~\bibnamefont{Schiffer}} \bibnamefont{and}
  \bibinfo{author}{\bibfnamefont{W.}~\bibnamefont{Rossow}},
  \bibinfo{journal}{Bull. Amer. Meteorol. Soc.} \textbf{\bibinfo{volume}{64}},
  \bibinfo{pages}{779} (\bibinfo{year}{1983}).

\bibitem[{\citenamefont{Freidenreich and Ramaswamy}(1993)}]{Freidenreich1993}
\bibinfo{author}{\bibfnamefont{S.~M.} \bibnamefont{Freidenreich}}
  \bibnamefont{and}
  \bibinfo{author}{\bibfnamefont{V.}~\bibnamefont{Ramaswamy}},
  \bibinfo{journal}{Journal of Geophysical Research: Atmospheres}
  \textbf{\bibinfo{volume}{98}}, \bibinfo{pages}{7255} (\bibinfo{year}{1993}),
  ISSN \bibinfo{issn}{2156-2202},
  \urlprefix\url{http://dx.doi.org/10.1029/92JD02887}.

\bibitem[{\citenamefont{Bernath}(2002)}]{Bernath2002}
\bibinfo{author}{\bibfnamefont{P.~F.} \bibnamefont{Bernath}},
  \bibinfo{journal}{Phys. Chem. Chem. Phys.} \textbf{\bibinfo{volume}{4}},
  \bibinfo{pages}{1501} (\bibinfo{year}{2002}),
  \urlprefix\url{http://dx.doi.org/10.1039/B200372D}.

\bibitem[{\citenamefont{Zheng and Ding}(2000)}]{Zheng2000}
\bibinfo{author}{\bibfnamefont{Y.}~\bibnamefont{Zheng}} \bibnamefont{and}
  \bibinfo{author}{\bibfnamefont{S.}~\bibnamefont{Ding}},
  \bibinfo{journal}{Journal of Molecular Spectroscopy}
  \textbf{\bibinfo{volume}{201}}, \bibinfo{pages}{109 } (\bibinfo{year}{2000}),
  ISSN \bibinfo{issn}{0022-2852},
  \urlprefix\url{http://www.sciencedirect.com/science/article/pii/S0022285200980603}.

\bibitem[{\citenamefont{Iachello and Levine}(1995)}]{Iachello1995}
\bibinfo{author}{\bibfnamefont{F.}~\bibnamefont{Iachello}} \bibnamefont{and}
  \bibinfo{author}{\bibfnamefont{R.~D.} \bibnamefont{Levine}},
  \emph{\bibinfo{title}{Algebraic Theory of Molecules}}
  (\bibinfo{publisher}{Oxford University Press, New York},
  \bibinfo{year}{1995}).

\end{thebibliography}

\end{document}